\documentclass[sigconf,screen]{acmart}
\AtBeginDocument{%
  }

\setcopyright{cc}
\setcctype{by}
\acmDOI{10.1145/3832783.3834496}
\acmYear{2026}
\copyrightyear{2026}
\acmISBN{979-8-4007-2882-2/2026/10}
\acmConference[ASE '26]{Proceedings of the 41st IEEE/ACM International Conference on Automated Software Engineering}{October 12--16, 2026}{Munich, Germany}
\acmBooktitle{Proceedings of the 41st IEEE/ACM International Conference on Automated Software Engineering (ASE '26), October 12--16, 2026, Munich, Germany}
\acmSubmissionID{ase26ind-p143-p}
\received{2026-04-30}
\received[accepted]{2026-07-01}

\usepackage{tikz}
\usetikzlibrary{arrows.meta,fit,positioning,calc}
\usepackage{amsmath}
\usepackage{microtype}
\usepackage{algpseudocode}
\usepackage[english]{babel}
\usepackage[utf8]{inputenc}
\usepackage{hyphenat}
\usepackage{algorithm}
\usepackage{enumitem}
\usepackage{listings}
\usepackage[dvipsnames]{xcolor}
\usepackage{balance}
\lstdefinelanguage{yaml}{
  keywords={true, false, null, yes, no},
  keywordstyle=\color{BrickRed}\bfseries,
  sensitive=false,
  comment=[l]{\#},
  commentstyle=\color{Gray}\itshape,
  stringstyle=\color{OliveGreen},
  morestring=[b]',
  morestring=[b]",
}
\usepackage{soul}
\usepackage{multicol}
\usepackage{bbding}
\usepackage{svg}
\usepackage{ifthen}
\usepackage{csquotes}
\usepackage{float}
\usepackage[T1]{fontenc}
\usepackage{mathrsfs}
\usepackage{pifont}
\usepackage{booktabs}
\usepackage{array}
\usepackage[caption=false]{subfig}

\newcommand{\gq}{{\sc GraphQLer}}
\newcommand{\pt}{\fontfamily{lmtt}\selectfont}

\definecolor{commentcolor}{RGB}{63, 127, 95}
\definecolor{functioncolor}{RGB}{0, 0, 128}
\definecolor{keywordcolor}{RGB}{128, 0, 0}
\definecolor{builtin}{RGB}{51, 51, 255}
\newcommand{\yes}{\Checkmark}
\newcommand{\no}{\XSolidBrush}
\begin{document}

\title[GraphQLer: Enhancing GraphQL Security with Context-Aware API Testing]{GraphQLer: Enhancing GraphQL Security \\ with Context-Aware API Testing}









\author{Omar Tsai}
\orcid{0000-0002-6335-3335}
\affiliation{%
  \institution{Simon Fraser University}
  \city{Vancouver}
  \country{Canada}
}
\email{omar@ztasecurity.com}

\author{Jianing Li}
\orcid{0009-0005-6436-1192}
\affiliation{%
  \institution{Simon Fraser University}
  \city{Vancouver}
  \country{Canada}
}
\email{leemailbox1@gmail.com}

\author{Tsz Tung Cheung}
\orcid{0009-0001-9950-5564}
\affiliation{%
  \institution{Simon Fraser University}
  \city{Vancouver}
  \country{Canada}
}
\email{ttcheung@ieee.org}

\author{Lejing Huang}
\orcid{0009-0008-6450-2209}
\affiliation{%
  \institution{Simon Fraser University}
  \city{Vancouver}
  \country{Canada}
}
\email{lejing.huang@outlook.com}

\author{Hao Zhu}
\orcid{0009-0005-4767-5958}
\affiliation{%
  \institution{Simon Fraser University}
  \city{Vancouver}
  \country{Canada}
}
\email{i@hao.soy}

\author{Jianrui Xiao}
\orcid{0009-0000-0613-9159}
\affiliation{%
  \institution{Simon Fraser University}
  \city{Vancouver}
  \country{Canada}
}
\email{xiaojianruijames@gmail.com}

\author{Iman Sharafaldin}
\orcid{0009-0009-1726-1729}
\affiliation{%
  \institution{Forward Security}
  \city{Vancouver}
  \country{Canada}
}
\email{i.sharafaldin@fwdsec.com}

\author{Mohammad A. Tayebi}
\correspondingauthor
\orcid{0009-0006-8689-9038}
\affiliation{%
  \institution{Simon Fraser University}
  \city{Vancouver}
  \country{Canada}
}
\email{tayebi@sfu.ca}

\renewcommand{\shortauthors}{O. Tsai, J. Li, T. T. Cheung, L. Huang, H. Zhu, J. Xiao, I. Sharafaldin, and M. A. Tayebi}

\begin{abstract}
\begin{sloppypar}
GraphQL APIs power production systems across financial services, e-commerce, and social platforms, yet their most critical access-control vulnerabilities---Insecure Direct Object Reference (IDOR), Use-After-Free (UAF), and state-dependent injection---routinely escape automated security testing. Industry-standard scanners (\textsf{ZAP}) and the leading research fuzzer (\textsf{EvoMaster}) test operations in isolation and cannot compose the multi-step sequences these flaws require.
We present \gq{}, an open-source automated security testing framework built for production GraphQL APIs. \gq{} constructs a typed dependency graph from live schema introspection and synthesizes \emph{vulnerability chains}---ordered operation sequences targeting specific flaw classes. Three strategies cover the critical attack surface: topological SCC-traversal for general reachability, cross-user IDOR replay for broken access control, and \mbox{CREATE$\to$DELETE$\to$READ} synthesis for UAF.
On a production financial API (\textit{FinServ}), \gq{} identified eight potential vulnerabilities---including denial-of-service vectors that exposed stack traces and sensitive implementation details---without prior documentation or authentication credentials. On a self-hosted \textsf{Saleor} instance pinned to the CVE-2022-39275 commit, \gq{} reproduced all four broken-access-control mutations cited in the security advisory. On the 11 public APIs of the coverage set, \gq{} achieves 85.52\% mean \textit{PositiveCoverage} versus 29.29\% (\textsf{EvoMaster}) and 21.80\% (\textsf{ZAP}); across the 21 evaluated APIs it detects all 5 confirmed IDOR endpoints, UAF behavior on two controlled schemas, and confirms XSS and SQLi on DVGA (an independent third-party oracle)---while all baselines detect zero chain-based vulnerabilities.
\end{sloppypar}
\end{abstract}



\maketitle
\section{Introduction}

GraphQL has emerged as a dominant API technology, adopted widely for its schema-driven, single-endpoint design that lets clients compose arbitrary queries over richly typed data graphs~\cite{GraphQlASystematicMappingStudy,brito2020rest}. This expressiveness, however, introduces security challenges largely absent from REST APIs. Unlike REST, where HTTP verbs and URL structure encode intent, GraphQL exposes all operations through a single \texttt{/graphql} endpoint with no built-in access-control enforcement. Mutations can silently depend on resources created by earlier operations, and the server resolves these dependencies at runtime---invisible to any tool that tests operations in isolation.

This \emph{stateful dependency structure} gives rise to three practically important vulnerability classes. \textbf{Insecure Direct Object Reference (IDOR)}: a resource created by one user can be read or modified by a second, unprivileged user who supplies only the resource ID. \textbf{Use-After-Free (UAF)}: a soft-deleted resource remains accessible after deletion because the backing store is never purged---the same free-then-dereference pattern as memory-level UAF, manifesting here as unauthorized data disclosure. \textbf{Injection}: SQL, OS command, and XSS payloads may require API state initialized by prior mutations---particularly on authenticated or multi-tenant APIs where valid resource IDs gate access. IDOR and UAF share a structural property that makes them invisible to existing tools: \emph{no single operation, executed in isolation, is sufficient to trigger the vulnerability}. Injection occupies a middle ground: payloads can sometimes be sent in isolation on simple APIs, but in practice---on authenticated or multi-tenant APIs---exploiting injection requires valid resource IDs materialized by prior mutations, state that single-operation tools never construct.

Existing tools fail to model this stateful structure. Black-box scanners (ZAP~\cite{zap-proxy-graphql}, BurpSuite~\cite{burpsuite}) vary parameters of individual operations and flag HTTP-level anomalies; they never compose multi-step sequences. EvoMaster~\cite{belhadi2022whitebox}---the only published GraphQL fuzzer we reproduce for comparison---applies evolutionary search over payloads but does not synthesize cross-user or post-delete chains. Other published GraphQL testing work, including reinforcement-learning~\cite{grahql_rl_testing} and LLM-assisted~\cite{prediql2025} approaches, likewise adapts payloads for individual operations rather than composing sequences. Static analyzers~\cite{taint_analysis_for_graph_apis,karlsson2020automatic,Vargas2018DeviationTA} detect structural anti-patterns but miss runtime access-control flaws. As a result, IDOR and UAF vulnerabilities in GraphQL APIs routinely escape automated testing.

We present \textbf{\gq{}}, an automated security testing framework that, to our knowledge, is the first to treat \emph{vulnerability chains}---ordered operation sequences designed to expose a specific flaw class---as first-class constructs in GraphQL testing. \gq{} operates in two phases. The \emph{compilation phase} parses the introspection schema and constructs a typed dependency graph linking objects, queries, and mutations. The \emph{testing phase} traverses this graph using three specialized chain strategies that synthesize and execute the sequences needed to expose IDOR, UAF, and injection vulnerabilities. An optional LLM-assisted resolver handles ambiguous mutation semantics; a multi-auth profile engine manages credentials for cross-user scenarios. \gq{} also exposes its full pipeline as a \emph{Model Context Protocol}~\cite{modelcontextprotocol} server---a tooling contribution enabling AI-assisted security workflows.

\begin{sloppypar}
\noindent\textbf{Industry Impact.} On the production-scale {\pt FinServ} API, \gq{} operated fully black-box and identified eight potential vulnerabilities, including at least two denial-of-service vectors exposing stack traces and implementation details. The findings were shared with the security team, demonstrating an automated evidence-collection workflow for industrial triage.
\end{sloppypar}

\noindent We make the following contributions:

\begin{itemize}[label=$\diamond$]
    \item We introduce \emph{vulnerability chains}---ordered GraphQL operation sequences---and three chain strategies (topological SCC-condensation, IDOR cross-user replay, UAF graph-edge synthesis) that detect multi-step vulnerabilities missed by every existing tool, including production systems carrying real security debt.

    \item We present \gq{}, an open-source tool\footnote{Source: \url{https://github.com/omar2535/GraphQLer}.} implementing all three strategies together with an injection suite covering nine vulnerability categories (three confirmed on purpose-built ground-truth APIs), an LLM-assisted dependency resolver, and a multi-auth profile engine.

    \item We evaluate \gq{} on 21 APIs (11 public, 5 platform, 2 known-vulnerable, 3 purpose-built ground-truth). On the 11 public APIs of the coverage set, \gq{} achieves 85.52\% mean \textit{PositiveCoverage}, versus 29.29\% for \textsf{EvoMaster} and 21.80\% for \textsf{ZAP}; on the ground-truth APIs it detects all 5 confirmed IDOR endpoints, UAF behavior on two controlled schemas, and three injection categories; and it independently validates injection detection on DVGA.

\end{itemize}

\section{Background \& Related Work}
\label{sec:background}

\subsection{GraphQL Fundamentals}

GraphQL~\cite{GraphQlASystematicMappingStudy} is a strongly-typed API language where queries, mutations, and subscriptions are exposed through a single \texttt{/graphql} endpoint. Introspection returns schema metadata at runtime and is \gq{}'s compilation input. Unlike REST, mutation intent is not encoded by HTTP verbs, so create/update/delete semantics must be inferred from names and descriptions. \gq{} uses this schema view to infer hard/soft dependencies and synthesize executable chains.

\subsection{Related Work}

\begin{sloppypar}
\noindent\textbf{GraphQL Testing.}
Existing GraphQL tools operate on single operations. Vargas et al.~\cite{Vargas2018DeviationTA} and Karlsson et al.~\cite{karlsson2020automatic} generate queries from the schema. Zetterlund et al.~\cite{Zetterlund_2022} replay production traces (white-box). Belhadi et al.~\cite{belhadi2022whitebox} extend EvoMaster~\cite{evomaster,Arcuri2021} with GraphQL plugins; their black-box mode is the only published fuzzer we reproduce for comparison. Hatfield-Dodds and Dygalo~\cite{Schemathesis} apply property-based testing to REST and GraphQL but with simple find-replace substitution. Commercial scanners (ZAP \cite{zap-proxy-graphql}, BurpSuite \cite{burpsuite}) target HTTP-level issues. Specialized tools (GraphCrawler \cite{GraphCrawler}, CrackQL \cite{CrackQL}, GraphQL-Cop \cite{graphql-cop}) probe individual operations. A reinforcement-learning approach \cite{grahql_rl_testing} adapts payloads without modeling cross-operation state. Recent LLM-assisted approaches improve GraphQL query generation \cite{ganesan2024llm} and fuzzing~\cite{prediql2025} but focus on data retrieval or single-operation coverage rather than security chain synthesis. BenGQL~\cite{bengql2025} provides a benchmarking framework for GraphQL testing tools; it covers coverage and performance metrics but does not address chain-based vulnerability detection. None synthesize multi-step sequences for cross-user or post-delete attacks. Furthermore, all existing tools are HTTP-only and cannot exercise GraphQL subscription operations; as a tooling contribution, \gq{} extends coverage to WebSocket-based subscriptions via its subscription engine (no RQ evaluation; reported as an implementation capability).
\end{sloppypar}

\noindent\textbf{Dependency-Aware REST Testing.}
RESTler \cite{restler}, foREST \cite{lin2022forest}, and RestTestGen \cite{resttestgen} infer producer/consumer relationships and sequence REST requests \cite{DependencyAwareWebTestGeneration}. GraphQL's single-endpoint model requires different algorithms; mutation action semantics must be inferred rather than read from HTTP methods. To our knowledge, \gq{} is the first system to apply dependency-aware chain generation to GraphQL security. Recent work has extended dependency-aware testing with coverage guidance~\cite{wuppie2025} and reinforcement learning to adaptively prioritize operation sequences~\cite{kim2023rl}.

\noindent\textbf{IDOR and UAF.}
IDOR is a common instance of OWASP A01: Broken Access Control~\cite{owasp-top10}. REST-focused work seeds IDs from prior responses \cite{intellgent-rest-api-data-fuzzing} but does not synthesize the resource-creation step. Use-After-Free has received little attention at the API layer. Existing fuzzers never place DELETE before READ in the same chain because no dependency edge connects them in a standard graph. Both gaps are addressed by \gq{}'s UAF and IDOR chain strategies. Automated detection of Broken Access Control (BAC) violations in REST APIs has been explored via traffic mining~\cite{bacalarm2025} and metamorphic testing for excessive data exposure~\cite{edefuzz2024}. These approaches operate on REST APIs using recorded traffic or OpenAPI specifications; they do not synthesize CREATE~$\to$~ACCESS chains from a live schema, which is \gq{}'s approach.

\section{Concepts of \gq{}}
\label{sec:methodology}

\begin{figure}[t]
  \centering
  \includegraphics[width=\linewidth]{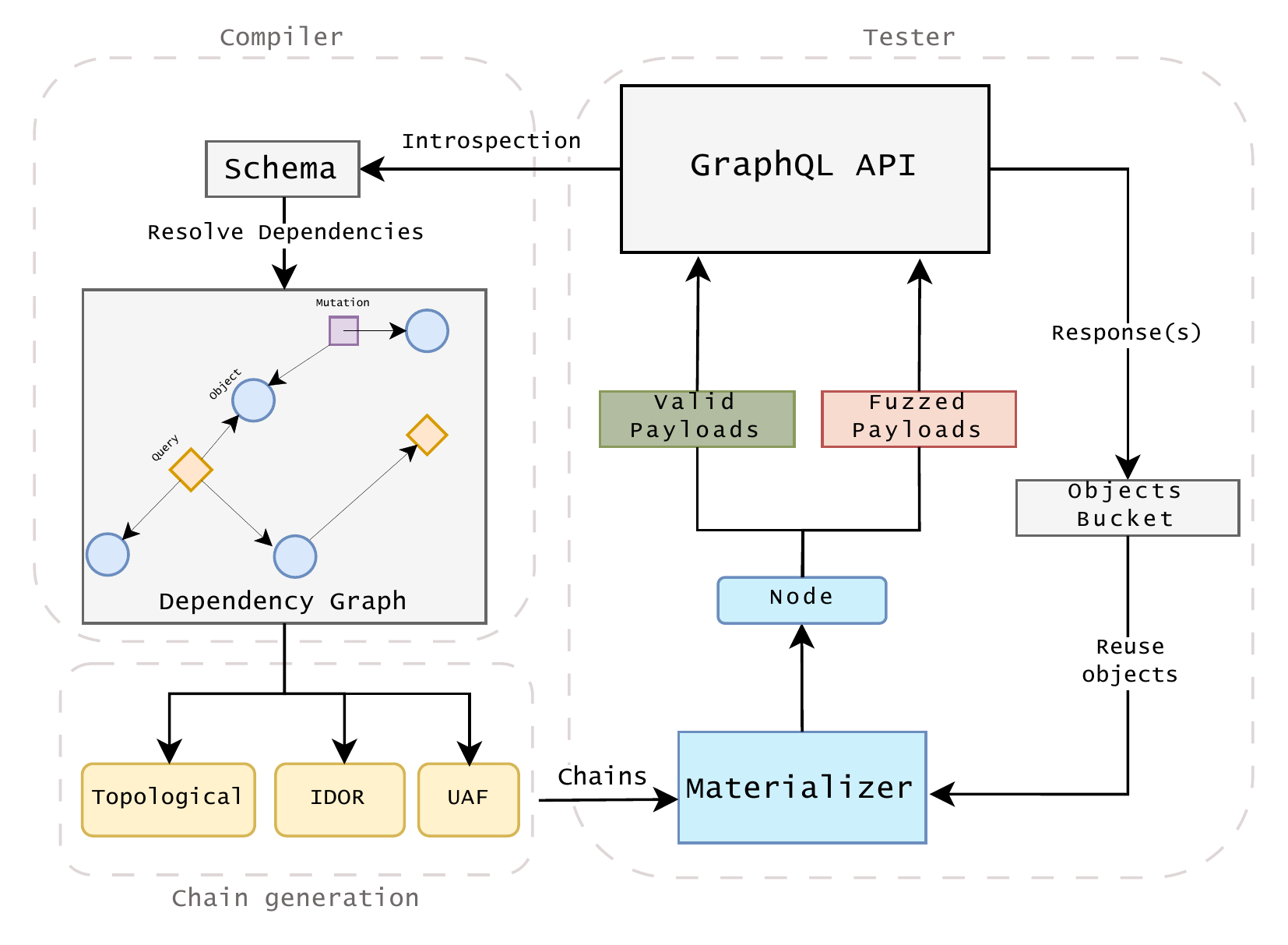}
  \caption{Workflow of \gq{}: the \emph{compilation} phase parses the live
    schema via introspection and constructs a typed dependency graph; the
    \emph{chain-generation} layer synthesizes ordered operation sequences
    using three strategies (topological SCC, IDOR cross-user replay, and
    UAF graph-edge synthesis); the \emph{testing} phase materializes
    payloads, dispatches requests to the API, and stores returned objects
    in the bucket for reuse when executing subsequent operations.}
  \Description{Three-phase workflow diagram: Compilation (schema introspection and dependency graph construction), Chain Generation (topological, IDOR, and UAF chain strategies), and Testing (materializer, objects bucket, valid and fuzzed payloads).}
  \label{fig:architecture}
\end{figure}

\gq{} operates in two phases (Figure~\ref{fig:architecture}). The \emph{compilation phase} parses the introspection schema (or falls back to word-list brute-forcing \cite{clairvoyance}) and builds a typed dependency graph. The \emph{testing phase} traverses this graph, materializes payloads, sends requests, and stores returned resources in an \emph{objects bucket} for reuse in subsequent operations.

\subsection{Compilation Phase}

The compilation phase builds the dependency graph in three steps:

\noindent {\bf Annotating Mutations with Actions.} GraphQL provides no explicit action semantics (unlike REST's HTTP verbs). The heuristic examines each mutation's name and description for action verbs and assigns \texttt{CREATE}, \texttt{UPDATE}, \texttt{DELETE}, or \texttt{UNKNOWN}. When the evidence is ambiguous or obfuscated, \texttt{UNKNOWN} avoids presenting an unsupported action as certain, but action-specific chains may not be constructed. A wrong known label can also change the three-pass traversal order: a missed \texttt{CREATE} can suppress an IDOR chain, a missed \texttt{DELETE} can suppress a UAF chain, and a wrong \texttt{UPDATE} label can move an operation into or out of the second pass before the intended object state is available. Thus annotation failures can cause false negatives. Conversely, an incorrect label can admit an inapplicable candidate, but the detector still requires response evidence rather than treating the label alone as a vulnerability. The optional LLM-based resolver uses full schema context to disambiguate cases such as \verb|destroyShip()|; its agreement and disagreement patterns are evaluated in Section~\ref{sec:llm-vs-heuristic}.

\noindent {\bf Inferring Dependencies.}
GraphQL explicitly defines object-to-object field relationships but does not expose query/mutation input dependencies in machine-readable form. We infer these missing edges by matching \texttt{ID}-typed input fields to object types in two stages. A \emph{name-pattern stage} first derives a candidate object name from the field: a suffixed field (\texttt{*Id}, \texttt{*Ids}) yields its stem, while a bare \texttt{id}/\texttt{ids} field is resolved from the operation's return type when available, falling back to the operation name with its CRUD prefix stripped. A \emph{similarity stage} then scores that candidate against the declared object types, retaining those whose normalized name is contained in the candidate and selecting the one with the smallest Levenshtein distance within a fixed threshold. When no object survives both stages the dependency is recorded as \texttt{UNKNOWN}; otherwise we draw a directed edge from the matched object to the operation. \texttt{NON-NULL} inputs produce \emph{hard dependencies} (\textit{hardDependsOn}); nullable inputs produce \emph{soft dependencies} (\textit{softDependsOn}). Three dependency types are resolved:

\begin{itemize}[leftmargin=*, label=$\diamond$, topsep=2pt, itemsep=0pt]
    \item \textit{Object-Object}: direct schema references; self-references require cycle-safe handling.
    \item \textit{Object-Mutation}: output edges from schema; input edges from ID-to-object matching.
    \item \textit{Object-Query}: bidirectional; output edges from schema, input edges by ID matching.
\end{itemize}

\begin{figure}[t!]
    \centering
    \includegraphics[width=\linewidth]{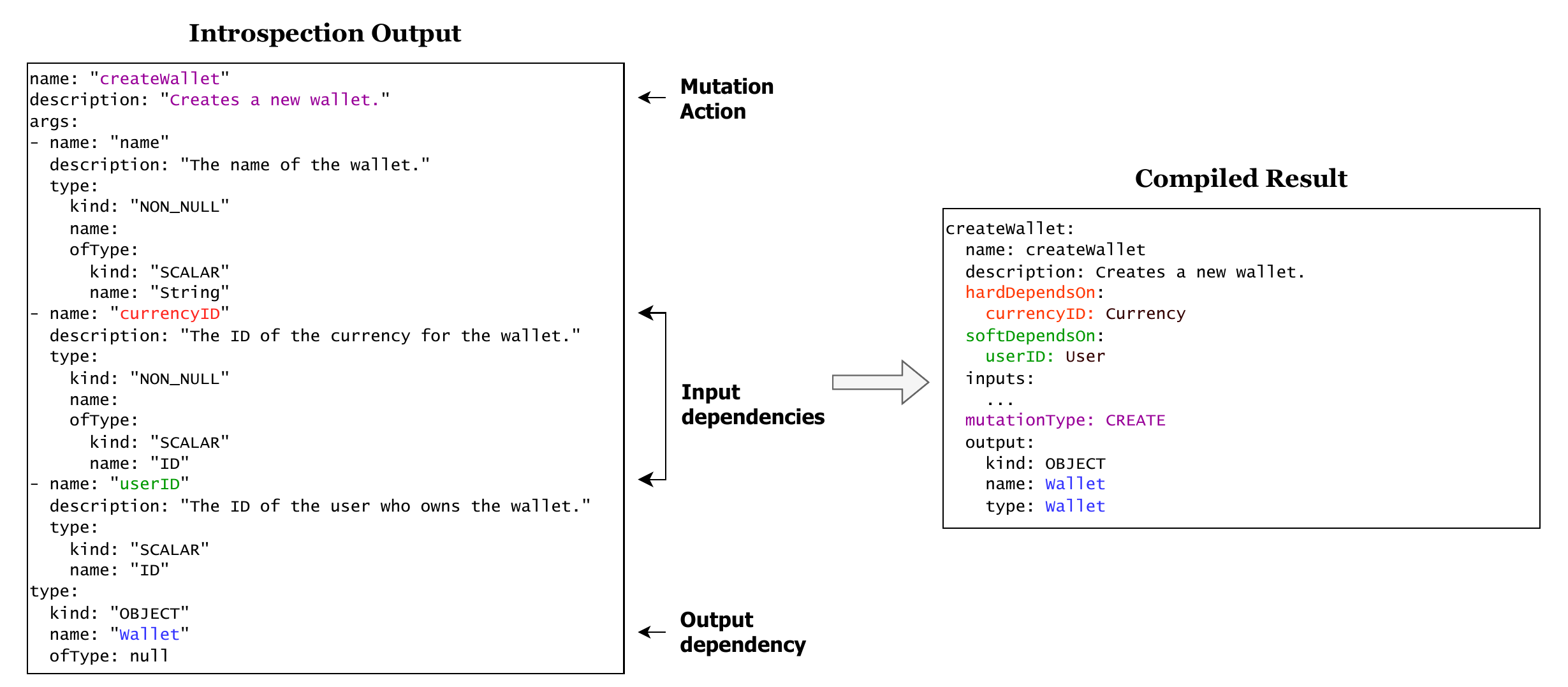}
    \caption{Compilation of a sample \texttt{createWallet} mutation.}
    \Description{Before/after annotated mutation node showing resolved hard- and soft-dependency edges to Currency and User objects, and a CREATE action label.}
    \label{fig:create_wallet_compiled}
\end{figure}

Figure~\ref{fig:create_wallet_compiled} illustrates the compiled result: \texttt{createWallet} is annotated \texttt{CREATE}, with a hard dependency on \texttt{Currency} (NON-NULL ID) and a soft dependency on \texttt{User} (nullable ID). Two inward edges come from those objects; one outward edge leads to the \texttt{Wallet} output.

\noindent {\bf Dependency Graph.} \gq{} creates one node per query, mutation, and object type, and adds directed edges per inferred dependency. Objects are not covered directly; traversal always passes through an operation (e.g., \texttt{User}~$\to$~\texttt{getWallet}~$\to$~\texttt{Wallet}).

\subsection{Testing Phase}
\begin{figure}[t!]
  \centering
  \includegraphics[width=80mm]{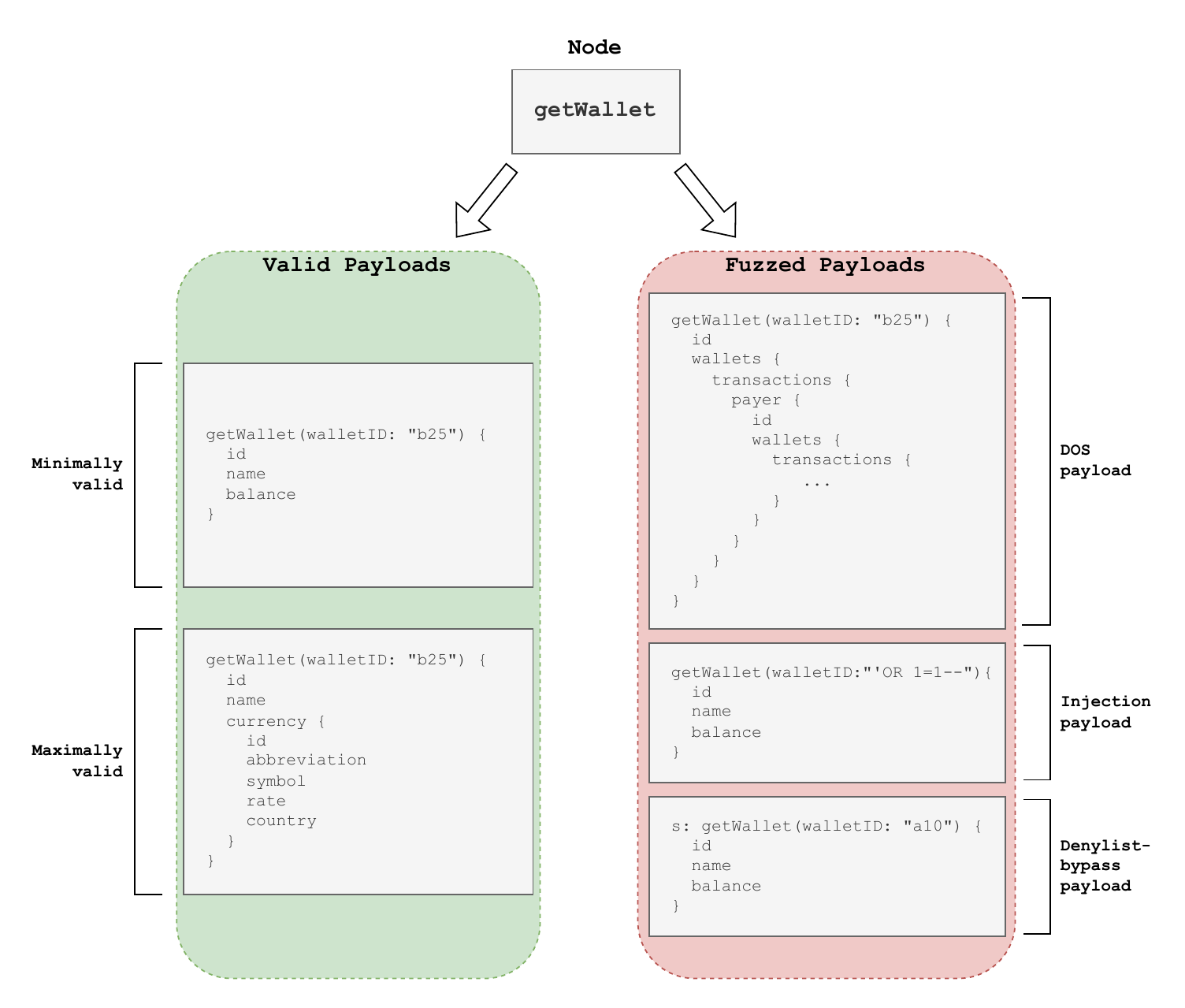}
  \caption{Payload generation for positive and negative testing.}
  \Description{Two-column diagram showing valid (minimally/maximally valid) and fuzzed (SQL injection, DoS, alias) payload variants generated by GraphQLer for each graph node.}
  \label{fig:payload-materialization}
\end{figure}

{\tolerance=9999\emergencystretch=20pt
The testing phase traverses the dependency graph, materializes payloads for each node, dispatches requests, and stores returned objects in the \emph{objects bucket} for reuse.

\noindent\textbf{Traversal.} \gq{} seeds traversal with all root nodes (in-degree zero); for fully cyclic graphs it relaxes this to minimum in-degree until at least one seed exists. Traversal uses multi-source DFS~\cite{multi-source-bfs} with a visited set to prevent re-visiting.

\noindent\textbf{Payload Materialization.} For each node, \gq{} generates two payload classes. \emph{Valid} payloads come in two variants: \emph{minimally valid} (fewest output selectors) and \emph{maximally valid} (all selectors). \emph{Fuzzed} payloads inject attack strings across nine categories---SQL injection, time-based blind SQL injection, NoSQL injection, XSS, OS command injection, SSRF, path traversal, HTML injection, and query-deny-list bypass---into input fields. For scalar inputs, \gq{} performs a name-matched lookup in the bucket; unresolved scalars receive random values. Hard-dependency ID fields abort materialization if absent from the bucket; soft-dependency IDs are skipped. Output fields are expanded recursively to materialize nested objects.

\noindent\textbf{Response Handling.} On success, all objects and scalars are stored in the bucket. On failure, \gq{} attempts to repair NON-NULL violations; otherwise, the node is deferred to the traversal stack tail and retried up to a configurable limit before being discarded.

\noindent\textbf{Three-Pass Traversal.} Graph traversal uses three ordered passes: \emph{(1)} \texttt{CREATE} mutations and queries; \emph{(2)} additionally \texttt{UPDATE} mutations; \emph{(3)} all remaining mutations including \texttt{DELETE}. All requests and responses are logged. Figure~\ref{fig:payload-materialization} illustrates the payload types generated at each node.
} 

\subsection{Vulnerability Chain Strategies}
\label{sec:chain-strategies}

While the three-pass traversal ensures broad operation coverage, the most security-critical vulnerabilities in GraphQL require \emph{ordered} sequences of operations---chains---where earlier steps create the preconditions for a later step to reveal a flaw. \gq{} generates vulnerability chains using three specialized strategies.

\noindent\textbf{Topological Chain Strategy.}
The dependency graph may contain cycles (e.g., a \texttt{getUser} query both requires and returns a \texttt{User} object). Naive DFS ordering breaks on such cycles. \gq{}'s topological chain strategy uses SCC-condensation (Algorithm~\ref{alg:topo}): each strongly-connected component (SCC) is collapsed into a single node, the resulting DAG is topologically sorted, and each SCC is then expanded in type-priority order (CREATE~$<$~Object~$<$~Query~$<$~UPDATE~$<$~DELETE). For every non-filtered node~$n$, the strategy computes all transitive ancestors, restricts to ancestors of the same filtered type-set, and builds a self-sufficient chain that can execute on an empty objects bucket.

\begin{algorithm}[t]
\footnotesize
\caption{Topological chain sort with SCC condensation}
\label{alg:topo}
\begin{algorithmic}[1]
\Require Dependency graph $G=(V,E)$
\Ensure Nodes in topological + type-priority order
\State $C \gets \textsc{Condensation}(G)$ \Comment{collapse SCCs into single nodes}
\State $order \gets \textsc{TopologicalSort}(C)$
\State $result \gets [\,]$
\ForAll{$c \in order$}
    \State $members \gets C[c].\text{nodes}$
    \State $result.\text{extend}(\textsc{SortByTypePriority}(members))$
\EndFor
\State \Return $result$
\end{algorithmic}
\end{algorithm}

\noindent\textbf{IDOR Chain Strategy.}
Insecure Direct Object Reference occurs when a resource created by one user (the \emph{primary} user) can be read, modified, or deleted by a second, unprivileged user (the \emph{secondary} user). \gq{}'s IDOR chain strategy takes the regular chains produced by the topological strategy as input and classifies each chain that contains a CREATE mutation followed by at least one additional operation. Qualifying chains are replayed with the CREATE step under the primary credential and all subsequent steps under the secondary credential. A detection is flagged when the secondary identity receives a non-empty, non-error response for a resource that should be private.

\noindent\textbf{UAF Chain Strategy.}
Use-After-Free (UAF) occurs when a resource remains readable after it has been deleted---a phantom-resource vulnerability. Unlike IDOR, the topological chain strategy \emph{never} places a DELETE mutation before a subsequent READ query in the same chain, because DELETE operations produce no output that the READ depends on, so no dependency edge exists between them. \gq{}'s UAF strategy compensates via graph-edge synthesis (Algorithm~\ref{alg:uaf}): it searches for quadruples $(c, o, d, r)$ where $c$ is a CREATE mutation producing object $o$, $d$ is a DELETE mutation consuming $o$, and $r$ is a Query consuming $o$. The synthesized chain is $c[\text{primary}] \to o[\text{primary}] \to d[\text{primary}] \to r[\text{post\_delete}]$. Here $o$ is the \emph{linking} element that carries the created resource's identity from $c$ to $d$ and $r$; consistent with Section~\ref{sec:methodology}, it is an object type rather than a dispatchable operation, so it contributes no request of its own and is not counted toward operation coverage. The three executable steps are $c$, $d$, and $r$. A detection is flagged when the post-delete step returns a non-null resource.

\begin{algorithm}[t]
\footnotesize
\caption{UAF chain synthesis from graph edges}
\label{alg:uaf}
\begin{algorithmic}[1]
\Require Dependency graph $G=(V,E)$
\Ensure Set of UAF candidate chains
\State $chains \gets [\,]$
\ForAll{$d \in V$ with $d.\text{mutationType} = \texttt{DELETE}$}
    \ForAll{$o \in \textsc{Predecessors}(G, d)$ with $o.\text{type} = \texttt{Object}$}
        \State $creates \gets \{c \in \textsc{Predecessors}(G, o) \mid c.\text{mutationType} = \texttt{CREATE}\}$
        \State $reads \gets \{r \in \textsc{Successors}(G, o) \mid r.\text{type} = \texttt{Query}\}$
        \ForAll{$(c, r) \in creates \times reads$}
            \State $\text{chain} \gets [c_{\text{primary}},\ o_{\text{primary}},\ d_{\text{primary}},\ r_{\text{post\_delete}}]$
            \State $chains.\text{append}(\text{chain})$
        \EndFor
    \EndFor
\EndFor
\State \Return $chains$
\end{algorithmic}
\end{algorithm}

\section{Implementation}
\label{sec:implementation}

\begin{sloppypar}
We have implemented \gq{} as a Python library and command-line tool (open source). The compilation and testing phases reside in the \texttt{compiler} and \texttt{fuzzer} modules, respectively. The dependency graph is persisted after compilation and loaded into memory at fuzzer start. All inputs and outputs are logged for inspection.

For authentication, \gq{} supports bearer or API tokens via command-line flags. For cross-user testing (IDOR), a second \texttt{--idor-auth} token activates the IDOR chain strategy. Multiple named auth profiles cover multi-tenant scenarios, and dedicated plugins handle refresh tokens or OpenID~Connect.
\end{sloppypar}

\noindent\textbf{Vulnerability Detection Suite.}
\gq{} integrates injection testing with its graph-based testing pipeline. During traversal, action labels determine the order in which mutations are exercised, and the \emph{objects bucket} carries returned IDs and state into subsequent requests. Fuzzed payloads can therefore reach injection points that depend on objects created or modified earlier in the sequence. The suite covers: (i)~\emph{SQL injection}, both error-based (pattern matching against MySQL, PostgreSQL, SQLite, and MSSQL error signatures) and time-based blind (detecting \texttt{pg\_sleep}/\texttt{SLEEP} delays); (ii)~\emph{NoSQL injection} with optional blind character-by-character data extraction; (iii)~\emph{OS command injection} (Unix/Windows payloads with output pattern matching); (iv)~\emph{XSS, path traversal, HTML injection, and SSRF}; and (v)~\emph{query-deny-list bypass}. \gq{} additionally performs \emph{ID enumeration} (probing sequential integer IDs with scope heuristics to suppress false positives on public endpoints) and \emph{field charset fuzzing} (detecting response-length variance from character set inputs). These detectors remain distinct from the IDOR and UAF strategies, but they operate over the same dependency-aware, stateful request sequences.

\noindent\textbf{LLM Integration (Optional).}
\gq{}'s LLM features are disabled by default; all RQ1--RQ3 experiments use heuristic-only mode to ensure reproducibility without API-cost dependencies. When enabled, the LLM is accessed via OpenAI-compatible APIs through \textsc{LiteLLM}~\cite{litellm} (default: \texttt{gpt-4o-mini}). It is invoked in two places: (i)~the \emph{dependency resolver} disambiguates mutation actions (e.g., \texttt{destroyShip}) that the heuristic labels \texttt{UNKNOWN}; (ii)~the \emph{chain classifier} re-scores IDOR/UAF candidates below $\tau=0.5$. A comparative evaluation of heuristic vs.\ LLM resolvers across 10 APIs appears in Section~\ref{sec:llm-vs-heuristic}.

\noindent\textbf{Subscription Support.}
\gq{} extends compilation and testing to GraphQL \emph{subscriptions}---real-time push operations defined alongside queries and mutations in the schema. The subscription parser extracts subscription definitions during the compilation phase, and a dedicated WebSocket-based materializer handles subscription payloads during testing, enabling coverage of event-driven API paths that are invisible to HTTP-only tools.

\noindent\textbf{Terminal User Interface (TUI).}
\gq{} provides an optional terminal user interface (\texttt{--tui} mode) built on the Textual framework~\cite{textual}. The TUI offers real-time log streaming, a chain explorer for inspecting synthesized chains, a query editor for manual payload crafting, and live statistics. This lets security practitioners inspect chains and craft payloads interactively.

\noindent\textbf{Model Context Protocol (MCP) Server.}
\gq{} exposes its compile, fuzz, and run pipeline as a \emph{Model Context Protocol}~\cite{modelcontextprotocol} (MCP) server (\texttt{--mcp} flag), enabling AI coding assistants (e.g., Claude, Cursor) to drive complete security testing workflows programmatically.

To prevent overloading target APIs, \gq{} enforces configurable per-request timeouts and inter-request delays. All requests and responses are logged and can be replayed using external tools such as Postman~\cite{postman} or Bruno~\cite{bruno} to ensure reproducibility.

\section{Evaluation}
\label{sec:exp_and_results}

We address three research questions:

\begin{itemize}[label=$\diamond$]
\item {\bf RQ1:} Does \gq{} achieve higher operation coverage (positive and negative) on GraphQL APIs compared to baseline methods?

\item {\bf RQ2:} Can \gq{} detect multi-step security vulnerabilities (IDOR, UAF, injection) that single-operation tools miss?

\item {\bf RQ3:} What is the individual contribution of each class-specific detection component (IDOR strategy, UAF strategy, injection suite) to vulnerability detection, and how does disabling each component affect the results?
\end{itemize}

\begin{table}[t] 
    \centering
  \footnotesize
    \caption{Public GraphQL APIs used in baseline testing (schema sizes at time of evaluation; live APIs may have grown).} 
    \label{tab:apis_tested} 
    \begin{tabular}{lccc} 
        \toprule 
        \textbf{API} & \textbf{\#Queries} & \textbf{\#Mutations} & \textbf{\#Objects} \\
        \midrule 

        {\pt User Wallet} \cite{UserWalletApi} & 11 & 15 & 8\\
        {\pt Food Delivery} \cite{Li_Sample_Food_Delivery_2023} & 5 & 5 & 8\\
        {\pt Countries} \cite{countries-api}  & 6 & 0 & 6\\
        {\pt Rick \& Morty} \cite{rick-morty} & 9 & 0 & 8 \\
        {\pt JSON-GraphQL} \cite{jsonGraphqlServer}  & 9 & 15 & 6\\
        {\pt GraphQLZero} \cite{GraphQLZero}  & 13 & 19 & 20\\
        {\pt AniList} \cite{aniListAPI} & 27 & 29 & 127 \\
        {\pt EHRI} \cite{ehriAPI}  & 19 & 0 & 46 \\
       {\pt Universe} \cite{universeAPI}  & 65 & 111 & 268 \\
        {\pt PokeAPI} \cite{PokeApi} & 459 & 0 & 1991\\
        {\pt TCGdex} \cite{tcgdex-api} & 6 & 0 & 14\\
        \bottomrule 
    \end{tabular}
\end{table}

\subsection{Experimental Design}
\subsubsection{Evaluated Applications} We evaluate \gq{} across four API categories: publicly available APIs, proprietary platform APIs, known-vulnerable applications, and purpose-built ground-truth APIs.

\begin{itemize}[leftmargin=*, label=$\diamond$]

\item \textit{Public APIs.} We utilize 11 public APIs across three categories: open-source, custom-built, and openly-hosted APIs \cite{public-graphql-apis, UserWalletApi, Li_Sample_Food_Delivery_2023}. For open-source APIs, we download the back-end code, set up a self-hosted GraphQL server, and run tests. For custom-built APIs, we develop GraphQL APIs that mirror the structure of publicly accessible server-built APIs. For openly-hosted APIs, we reference free-to-use online APIs on GitHub and conduct direct testing. The list of APIs is shown in Table \ref{tab:apis_tested}.

\item \textit{Platform APIs.} Five production GraphQL APIs operated by third parties, including GitLab~\cite{gitlab} and Yelp~\cite{yelp}, and the financial API we refer to throughout by the pseudonym {\pt FinServ}. Unlike the public set, these are live services we do not control; their schemas are not redistributable and some are covered by disclosure agreements, so we report aggregate detection counts (Section~\ref{sec:vuln-detection}) rather than per-API coverage.

\item \textit{Vulnerable Software.} We test against two known-vulnerable applications:
\begin{itemize}[leftmargin=*, label=$\triangleright$]

\item {\pt Saleor.} Saleor is an open-source, self-hosted platform that also offers a commercial version \cite{saleor}. The {\pt Saleor} API was found to be vulnerable to CVE-2022-39275 \cite{cve-2022-39275}, which exposes broken access control through several GraphQL mutations. In testing the {\pt Saleor} API, we aim to use previously reported CVEs associated with GraphQL APIs, focusing specifically on finding a reproducible CVE.

\item {\pt DVGA} (Damn Vulnerable GraphQL Application)~\cite{dvga} is a publicly available, independently designed vulnerable GraphQL application maintained by the security community. It contains documented injection vulnerabilities (SQL injection, XSS, OS command injection, SSRF, path traversal) and denial-of-service scenarios. Unlike the three purpose-built ground-truth APIs, DVGA was designed entirely independently of this research and serves as an independent oracle for validating \gq{}'s injection detection on a third-party API. DVGA has 12~queries, 7~mutations, and 14~object~types.

\end{itemize}

\item \textit{Ground-Truth Vulnerable APIs.}
To evaluate chain-based vulnerability detection with known ground truth, we use three purpose-built vulnerable APIs:

\begin{itemize}[leftmargin=*, label=$\triangleright$]

\item {\pt IDOR-API}. A deliberately vulnerable note-and-order service with two users: \emph{Alice} (owner) and \emph{Bob} (attacker). Five operations are IDOR-vulnerable: \texttt{getNote}, \texttt{updateNote}, \texttt{deleteNote}, \texttt{getOrder}, \texttt{deleteOrder}. The API has 4~queries, 5~mutations, and 6~object~types.

\item {\pt Injection-API}. A Node.js API with five confirmed injection instances across four operations: SQL injection (\texttt{searchPosts}), stored XSS (\texttt{createPost}), and both path traversal and reflected XSS (\texttt{readFile}), and OS command injection (\texttt{executeCommand}). The API has 3 queries, 2 mutations, and 5 object types.

\item {\pt VVSM-API} (Very Vulnerable Social Media). A FastAPI service with UAF (\texttt{deletePost} soft-deletes; \texttt{getPost} returns full data for deleted posts, HTTP~200), stored XSS, and SQLi; 4 queries, 4 mutations, 8 object types.

\end{itemize}

\end{itemize}

\subsubsection{Baselines} We compare our solution against the following baselines:

{\sloppy
\begin{itemize}[leftmargin=*, label=$\diamond$]
    \item \textsf{GenericTester}: \gq{} with no dependency graph and no object cache---a baseline that sends random syntactically valid payloads.

    \item \textsf{EvoMaster}: The only published GraphQL black-box fuzzer we reproduce for comparison~\cite{belhadi2022whitebox}. Dynamically varies payloads via evolutionary search over a 3{,}600\,s session budget.

    \item \textsf{ZAP}: OWASP's open-source black-box scanner~\cite{zap-proxy-graphql} with GraphQL API testing support.

    \item \textsf{BurpSuite}: A commercial web security proxy~\cite{burpsuite}; excluded from both RQ1 and RQ2 in this evaluation as it cannot be run non-interactively.

    \item \textsf{Schemathesis}: A property-based testing tool~\cite{Schemathesis} that generates and validates GraphQL queries against the schema; does not perform injection testing or chain synthesis.

    \item {\textsf{GraphQL-Cop}}~\cite{graphql-cop}: A GraphQL-specific security scanner included in the DVGA independent-oracle comparison; probes for misconfiguration and information-disclosure issues.

\end{itemize}
}

\noindent For the component ablation (Section~\ref{sec:comparison_baseline}), we introduce two variants of the proposed solution.

\begin{itemize}[leftmargin=*, label=$\diamond$]
    \item \textsf{GraphQLerOOB}: \gq{} without the dependency graph (objects bucket only).
    \item \textsf{GraphQLerODG}: \gq{} without the objects bucket (dependency graph only).

\end{itemize}

\subsubsection{Evaluation Measures} To address \textbf{RQ1}, we define a coverage metric for black-box testing that includes both positive and negative testing scenarios, as follows:

$$ \mathbf{\textit{PositiveCoverage}} = \frac{\#\text{NoErrors}}{\#\text{Endpoints}} $$

$$ \mathbf{\textit{NegativeCoverage}} = \frac{\#\text{Errors}}{\#\text{Endpoints}} $$

\begin{table}[t]
\footnotesize
\caption{Tool versions and configuration used in evaluation.}
\label{tab:tool_config}
\centering
\begin{tabular}{lll}
\toprule
\textbf{Tool} & \textbf{Version} & \textbf{Key settings} \\
\midrule
\gq{} & 2.3.8 & default; LLM disabled; 3{,}600\,s timeout \\
\textsf{ZAP} & 2.17.0 & GraphQL scanner enabled; active scan \\
\textsf{Schemathesis} & 4.15.2 & GraphQL mode; 100 examples per operation \\
\textsf{EvoMaster} & 3.2.0 & black-box mode; 3{,}600\,s budget \\
\textsf{GraphQL-Cop} & 1.15 & default; JSON output mode \\
\textsf{GenericTester} & --- & \gq{} with graph and bucket disabled \\
\bottomrule
\end{tabular}
\end{table}

\noindent where \textit{PositiveCoverage} is the proportion of endpoints that return at least one successful response containing data (HTTP 2xx with a non-null \texttt{data} field and no GraphQL-layer \texttt{errors} field), and \textit{NegativeCoverage} is the proportion of endpoints that produce errors despite valid inputs (HTTP 4xx/5xx or a GraphQL \texttt{errors} field in an HTTP 200 response). These are the only observable black-box signals when source code is inaccessible---the same proxy metrics used in REST API testing~\cite{restler,resttestgen}. Reachability is a necessary precondition for security testing: an endpoint that a tool cannot reach with a valid response cannot be subjected to vulnerability-triggering payloads or chain preconditions. Positive coverage therefore bounds the set of operations on which any security finding is possible. We acknowledge that this metric is structurally aligned with \gq{}'s core mechanism---dependency-aware traversal is designed to maximize reachable endpoints---and should be interpreted alongside the ground-truth detection results (Section~\ref{sec:chain-vuln-eval}), where vulnerability findings on independently validated APIs provide a complementary measure of security effectiveness. For ground-truth APIs, we supplement coverage with precision and recall over confirmed vulnerabilities.

\begin{table*}
\footnotesize
\caption{Mean positive (+) and negative (-) operation coverage. \texttt{FAILED} denotes no usable endpoint-level output.}
    \centering
    \setlength{\tabcolsep}{5.5pt} 
    \renewcommand{\arraystretch}{1.1}
    \begin{tabular}{lcccccccccccc}
        \toprule
      \textbf{API} & \multicolumn{2}{c}{\textsc{GenericTester}} & \multicolumn{2}{c}{\textsc{ZAP}} & \multicolumn{2}{c}{\textsc{Schemathesis}} & \multicolumn{2}{c}{\textsc{GraphQL-Cop}} & \multicolumn{2}{c}{\textsc{EvoMaster}} & \multicolumn{2}{c}{\textbf{\gq{}}} \\
        \cmidrule(lr){2-3} \cmidrule(lr){4-5} \cmidrule(lr){6-7} \cmidrule(lr){8-9} \cmidrule(lr){10-11} \cmidrule(lr){12-13}
      & \textbf{(+)} & \textbf{(-)} & \textbf{(+)} & \textbf{(-)} & \textbf{(+)} & \textbf{(-)} & \textbf{(+)} & \textbf{(-)} & \textbf{(+)} & \textbf{(-)} & \textbf{(+)} & \textbf{(-)} \\
        \midrule
      {\pt User Wallet} \cite{UserWalletApi}         & 30.77\% & 3.85\% & 15.40\% & 26.90\% & \texttt{N/A} & {\bf 28.80\%} & \texttt{N/A} & 3.85\% & 25.00\% & 0.00\% & {\bf 100\%} & 16.92\% \\
      {\pt Food delivery} \cite{Li_Sample_Food_Delivery_2023}       & 30.00\% & 0.00\% & 20.00\% & {\bf 50.00\%} & \texttt{N/A} & 40.00\% & \texttt{N/A} & 10.00\% & 37.00\% & 0.00\% & {\bf 100\%} & 30.00\% \\
      {\pt Countries} \cite{countries-api}          & 50.00\% & 50.00\% & 0.00\% & 0.00\% & \texttt{N/A} & 25.00\% & \texttt{N/A} & 16.67\% & 26.67\% & 40.00\% & {\bf 100\%} & {\bf 68.75\%} \\
      {\pt Rick \& Morty} \cite{rick-morty}         & 33.33\% & 0.00\% & 20.00\% & 0.00\% & \texttt{N/A} & 22.20\% & \texttt{N/A} & 11.11\% & 68.89\% & 0.00\% & {\bf 97.78\%} & {\bf 64.44\%} \\
      {\pt JSON-GraphQL} \cite{jsonGraphqlServer} & 41.67\% & 0.00\% & 14.29\% & 28.57\% & \texttt{N/A} & {\bf 37.10\%} & \texttt{N/A} & 0.00\% & 33.33\% & 0.00\% & {\bf 100\%} & 12.08\% \\
      {\pt GraphQLZero} \cite{GraphQLZero}        & 37.50\% & 6.25\% & 37.50\% & 37.50\% & \texttt{N/A} & {\bf 86.20\%} & \texttt{N/A} & 3.12\% & 49.06\% & 7.50\% & {\bf 75.00\%} & 21.88\% \\
      {\pt AniList} \cite{aniListAPI}             & 3.93\% & 96.07\% & 0.00\% & 0.00\% & \texttt{N/A} & 44.30\% & \texttt{N/A} & 0.54\% & 0.00\% & 0.00\% & \textbf{92.86\%} & \textbf{99.33\%} \\
      {\pt EHRI} \cite{ehriAPI}                & 42.11\% & 10.53\% & 0.00\% & 0.00\% & \texttt{N/A} & 24.00\% & \texttt{N/A} & 5.26\% & 18.42\% & 0.00\% & {\bf 100\%} & {\bf 57.89\%} \\
      {\pt Universe} \cite{universeAPI}            & 18.75\% & 63.64\% & 6.20\% & {\bf 92.60\%} & \texttt{N/A} & 86.10\% & \texttt{N/A} & 0.51\% & 23.86\% & 2.39\% & \textbf{43.69\%} & 34.79\% \\
      {\pt PokeAPI} \cite{PokeApi}             & 11.20\% & {\bf 67.29\%} & 26.40\% & 0.00\% & \texttt{FAILED} & \texttt{FAILED} & \texttt{N/A} & 0.22\% & 0.00\% & 0.00\% & {\bf 31.37\%} & 24.84\% \\
      {\pt TCGdex} \cite{tcgdex-api}              & 50.00\% & 50.00\% & {\bf 100\%} & 0.00\% & \texttt{N/A} & \textbf{100\%} & \texttt{N/A} & 16.67\% & 40.00\% & 0.00\% & \textbf{100\%} & 50.00\% \\
        \bottomrule
    \end{tabular}

    \label{tab:coverage_table}
\end{table*}

All tests are conducted on a server equipped with 16 CPU cores, 64\,GB of RAM, and Ubuntu 22.04. Table~\ref{tab:tool_config} lists the tool versions and key configuration parameters used; all tools ran with default settings except where noted.

Each API--tool pair was scheduled for ten runs. Table~\ref{tab:coverage_table} averages completed runs; \texttt{FAILED} denotes no usable output. The archived \textsf{ZAP} run for JSON-GraphQL used the earlier 21-operation schema; its other entries and the remaining tools use the listed snapshots. Raw outputs and completion counts are archived.

\subsection{Test Coverage}
\label{sec:comparison_baseline}

\gq{} achieves a mean \textit{PositiveCoverage} of 85.52\%, versus 29.29\% for \textsf{EvoMaster} and 21.80\% for \textsf{ZAP} (absolute gains of 56.22 and 63.72\,pp), leading on 11/11 and 10/11 APIs respectively. The gap is steepest on mutation-heavy schemas---on {\pt User Wallet} and {\pt Food Delivery}, \gq{} reaches 100\% while \textsf{EvoMaster} stalls at 25\% and 37\%, its successes confined to mutations that take freshly constructed inputs. Both baselines report 0\% on {\pt AniList}; \textsf{EvoMaster} also fails on {\pt PokeAPI}, where the 459-operation schema exhausts its search budget. \textsf{Schemathesis} and \textsf{GraphQL-Cop} are excluded from the aggregate and marked \texttt{N/A} for \textit{(+)}: \textsf{Schemathesis} validates schema conformance rather than exercising operations for data, and \textsf{GraphQL-Cop} issues a fixed set of security probes rather than traversing schema operations.

For \textit{NegativeCoverage}, \gq{} leads \textsf{EvoMaster} on all 11 APIs and \textsf{ZAP} on 6/11. The five ZAP losses ({\pt User Wallet}, {\pt Food Delivery}, {\pt GraphQLZero}, {\pt Universe}, {\pt JSON-GraphQL}) occur on APIs with strict validation, where \textsf{ZAP}'s aggressive mutation readily triggers errors that \gq{}'s schema-constrained payloads avoid---a deliberate trade-off: valid payloads are a prerequisite for chain construction. \textsf{GraphQL-Cop}'s negative coverage is uniformly low (0.00\%--16.67\%), as expected for a scanner whose probes target only a handful of misconfiguration checks. Overall, Table~\ref{tab:coverage_table} supports {\bf RQ1}: \gq{} provides substantially broader positive coverage than all baselines while also leading \textsf{EvoMaster} on negative coverage throughout.

\subsubsection{Ablation Study}

\begin{sloppypar}
Combining the dependency graph with the objects bucket increases mean \textit{PositiveCoverage} by 27.4 percentage points over \textsf{GraphQLerODG} (graph only) and 22.7 over \textsf{GraphQLerOOB} (bucket only), confirming both components are jointly necessary. To keep this sweep tractable, it uses a reduced 60\,s per-API budget with maximal payloads disabled, applied identically to all three configurations; absolute values are therefore lower than in Table~\ref{tab:coverage_table} and should be compared only within the ablation.
\end{sloppypar}

\subsection{Vulnerability Detection}
\label{sec:vuln-detection}
This section compares \gq{} and baselines on vulnerability detection across public APIs, platform APIs, known-vulnerable applications, and purpose-built ground-truth APIs.

\noindent\textbf{Public \& Platform APIs.} \gq{} flags potential vulnerabilities from HTTP~500 responses (server-side failures) and HTTP~200 responses with embedded error messages or unexpected data disclosures.

For reproducible quantitative comparison, we report only automatable tools (\textsf{ZAP}, \textsf{EvoMaster}, \textsf{Schemathesis}). \textsf{BurpSuite} is intentionally omitted from quantitative results because its workflow is interactive and cannot be executed in a fully automated, repeatable batch setup under the same harness.

\begin{sloppypar}
Across the 16 public and platform APIs, \gq{} detected 11 potential vulnerabilities. These are alerts requiring analyst triage rather than confirmed exploits.
\end{sloppypar}

\noindent\textbf{FinServ.} \gq{} identified eight \emph{potential} vulnerabilities in {\pt FinServ}, including at least two denial-of-service vectors that exposed stack traces and sensitive implementation details. These automated findings require analyst triage; the available evidence consisted of response behavior and error content observed during black-box testing. Under the research agreement, the findings were shared with the FinServ security team before submission for confirmation, impact assessment, and remediation. The subsequent confirmation and remediation status are outside the scope of this evaluation. In practice, this makes \gq{} a prioritization and evidence-collection tool for security teams, with human analysts deciding whether a finding warrants disclosure, a patch, or additional controlled reproduction.

\noindent\textbf{Saleor.} On a self-hosted {\pt Saleor} instance pinned to the CVE-2022-39275 commit, \gq{} identified all four vulnerable mutations (broken access control), reproducing the exact error codes in the security advisory.

Competing automatable tools also exhibited reliability failures (\textsf{ZAP} crashed on Yelp; \textsf{EvoMaster} hit a Java heap error on GitLab), while \gq{} produced usable output for every target.

\subsection{Chain-Based Vulnerability Detection (RQ2)}
\label{sec:chain-vuln-eval}

To evaluate \gq{}'s chain-based detection capabilities with known ground truth, we run \gq{} and the compared baselines (\textsf{ZAP}, \textsf{Schemathesis}, \textsf{EvoMaster}, \textsf{GraphQL-Cop}) on the three purpose-built vulnerable APIs. None of the baselines detect any chain-based vulnerabilities (IDOR or UAF). This result establishes a structural capability gap rather than a performance difference within a shared capability class: single-operation tools are architecturally incapable of constructing multi-step chains, so their zero detection rate is a definitional consequence of their design---not an empirical shortfall. \gq{} fills this structural gap; VVSM-API provides the formal UAF ground truth.

\begin{sloppypar}
\noindent\textbf{IDOR Detection.} \gq{} runs on {\pt IDOR-API} with Alice's (primary) and Bob's (secondary) tokens, achieving 100\% operation coverage (9/9 endpoints) in approximately 6.8\,s (the small schema size means the full traversal completes well within the 3{,}600\,s session cap). \gq{} correctly identifies all 5 ground-truth vulnerable operations: \texttt{getNote}, \texttt{updateNote}, \texttt{deleteNote}, \texttt{getOrder}, and \texttt{deleteOrder}. Chains such as \texttt{createNote}~[primary]~$\to$~\texttt{getNote}~[secondary] confirm Bob receives Alice's resource; \texttt{deleteNote}~[secondary] returning \texttt{\{``ok'':~true\}} confirms delete-based IDOR. 
\end{sloppypar}

\begin{sloppypar}
\noindent\textbf{UAF Detection.} We run \gq{} on {\pt VVSM-API} with Alice's token. \gq{} achieves 87.5\% standalone operation coverage (7/8 endpoints) in approximately 3.1\,s; the remaining endpoint requires object state the bucket cannot resolve independently. The UAF chain strategy synthesizes the quadruple (\texttt{createPost}, \texttt{Post}, \texttt{deletePost}, \texttt{getPost}) via Algorithm~\ref{alg:uaf}: after \texttt{deletePost} executes, \texttt{getPost} still returns the deleted resource's full data (HTTP~200). The topological strategy alone produces no chain containing both DELETE and the subsequent READ, confirming graph-edge synthesis is essential. \gq{} additionally confirms stored XSS on \texttt{createPost} and SQL injection on \texttt{createComment}.
\end{sloppypar}

\begin{table}[t]
\footnotesize
\caption{Ground-truth detection results, grouped by API. \yes: Detected; \no: Not detected; $\sim$: Potential. \textit{Schem.}: Schemathesis; \textit{EvoM.}: EvoMaster; \textit{Cop}: GraphQL-Cop.}
\label{tab:chain_vuln_results}
\centering
\setlength{\tabcolsep}{3pt}
\begin{tabular}{lccccc}
\toprule
\textbf{Vulnerability} & \textbf{ZAP} & \textbf{Schem.} & \textbf{EvoM.} & \textbf{Cop} & \textbf{\gq{}} \\
\midrule
\multicolumn{6}{@{}l}{\textit{IDOR-API}} \\
IDOR (\texttt{getNote})    & \no & \no & \no & \no & \yes \\
IDOR (\texttt{updateNote}) & \no & \no & \no & \no & \yes \\
IDOR (\texttt{getOrder})   & \no & \no & \no & \no & \yes \\
IDOR (\texttt{deleteNote}) & \no & \no & \no & \no & \yes \\
IDOR (\texttt{deleteOrder})& \no & \no & \no & \no & \yes \\
\midrule
\multicolumn{6}{@{}l}{\textit{Injection-API}} \\
SQL Injection              & \yes & \no & \no & \no & \yes \\
OS Command Inj.            & \yes & \no & \no & \no & \yes \\
XSS (2 operations)         & \no & \no & \no & \no & \yes \\
Path Traversal             & \no & \no & \no & \no & $\sim$ \\
\midrule
\multicolumn{6}{@{}l}{\textit{VVSM-API}} \\
UAF (\texttt{getPost})     & \no & \no & \no & \no & \yes \\
XSS (\texttt{createPost})  & \no & \no & \no & \no & \yes \\
SQL Injection              & \no & \no & \no & \no & \yes \\
\midrule
\multicolumn{6}{@{}l}{\textit{DVGA}} \\
XSS (\texttt{audits})      & \no & \no\textsuperscript{$\dagger$} & \no & \no & \yes \\
SQL Injection (\texttt{pastes}, \texttt{audits}) & \no & \no\textsuperscript{$\dagger$} & \no & \no & \yes \\
\bottomrule
\multicolumn{6}{@{}p{0.97\columnwidth}@{}}{\footnotesize \textsuperscript{$\dagger$}Schemathesis completed on DVGA (exit code 2 from non-fatal errors) and reported no confirmed injection findings.}
\end{tabular}
\end{table}

\noindent\textbf{Injection Detection.} We run \gq{} on {\pt Injection-API} without authentication. \gq{} achieves 100\% operation coverage (5/5 endpoints) in approximately 1.4\,s and confirms three injection vulnerability categories: SQL injection (\texttt{searchPosts}; SQLite error pattern), OS command injection (\texttt{executeCommand}; \texttt{/etc/passwd} content returned), and XSS on two operations---stored XSS on \texttt{createPost} and reflected XSS on \texttt{readFile}, both echoing the injected script tag verbatim in the response body. Path traversal payloads are reflected but file-content leakage is not automatically confirmed---a known limitation discussed in Section~\ref{sec:discussion}. \textsf{Schemathesis} generates 620--900 schema-conformant test cases per ground-truth API but raises no security alerts, confirming it tests conformance rather than vulnerability exploitation. \textsf{ZAP} raises warning-level alerts for SQL injection and OS command injection but misses XSS and path traversal.

\begin{sloppypar}
\noindent\textbf{Independent Oracle Validation (DVGA).}
To address the limitation that all three purpose-built ground-truth APIs were designed by the research team, we additionally evaluate \gq{} on {\pt DVGA}~\cite{dvga} (Damn Vulnerable GraphQL Application)---an independently designed, community-maintained vulnerable GraphQL application with a publicly documented vulnerability list. \gq{} achieves 68.4\% operation coverage (13/19 endpoints) and confirms two injection vulnerability categories: stored XSS (\texttt{audits}; script tags reflected verbatim) and SQL injection (\texttt{pastes} and \texttt{audits}; database error patterns matched---\texttt{sqlite3} and \texttt{@@version}, respectively). \textsf{ZAP} produces 134 alerts (all generic web-server header configuration issues: missing CSP headers, cookie flags, content-type options) and detects zero injection vulnerabilities. \textsf{GraphQL-Cop} reports 8 positive checks out of 12, primarily GraphQL hardening and DoS-style findings (e.g., introspection and query-overloading), but no confirmed injection findings. \textsf{Schemathesis} completes (exit code~2 from non-fatal errors) and reports no confirmed injection findings. The 6 uncovered DVGA operations are all mutations and fall into two categories: (i)~four require authentication credentials not configured in this run (\texttt{login}, \texttt{createUser}, \texttt{importPaste}, \texttt{uploadPaste}), matching the auth-gap limitation documented in Section~\ref{sec:discussion}; (ii)~\texttt{deletePaste} and \texttt{editPaste} take a bare \texttt{Int} paste identifier that carries no schema-level type information, so no dependency edge links them to \texttt{createPaste} and the traversal never materializes a valid target---an instance of the scalar-materialization limitation in Section~\ref{sec:discussion}. Separately, \texttt{systemDiagnostics} is reached but its documented OS command injection is not confirmed: \texttt{cmd} executes only after valid \texttt{username}/\texttt{password} arguments, a multi-field precondition that \gq{}'s current injection engine does not yet model as a chain precondition---a concrete future-work direction. These results confirm that \gq{}'s injection detection generalizes to an independently-designed API that the research team did not instrument.
\end{sloppypar}

\noindent\textbf{Precision and False Positives.} \gq{} raises 11 IDOR chain alerts on IDOR-API. All 5 ground-truth vulnerable operations are among them, giving a recall of 5/5. Of the remaining 6, four are raised on object-type nodes, which are not dispatchable operations (Section~\ref{sec:chain-strategies}); they emit HTTP~0 and are automatically filterable. The other two (\texttt{myNotes}, \texttt{myOrders}) do expose Alice's data to Bob, but whether the listing semantics constitute a flaw requires analyst triage, so we score them conservatively as false positives. Raw precision is therefore 5/11 (45.5\%), rising to 5/7 (71.4\%) once object-type alerts are filtered. On Injection-API and VVSM-API no chain alert is raised on an operation outside the ground-truth set.

\gq{} detects all three vulnerability classes across three purpose-built APIs (Table~\ref{tab:chain_vuln_results}). VVSM-API provides the formal UAF oracle; IDOR-API also exhibits post-delete resource access (Section~\ref{sec:chain-ablation}), supplying within-suite cross-schema evidence. Independent validation of IDOR and UAF on an external application remains future work. Chain-based vulnerabilities (IDOR and UAF) are uniquely identified by \gq{} and missed entirely by all baselines, fully addressing {\bf RQ2} within these controlled benchmarks.

\subsection{Detection-Component Ablation (RQ3)}
\label{sec:chain-ablation}

\begin{sloppypar}
\raggedright
To quantify the contribution of each class-specific detection component, we run three ablated variants of \gq{} on the ground-truth APIs established in the previous section: \textsf{\gq{}-noIDOR} (IDOR disabled), \textsf{\gq{}-noUAF} (UAF disabled), and \textsf{\gq{}-noInj} (injection disabled). Table~\ref{tab:chain_ablation} reports detection recall for each variant.
\end{sloppypar}

\begin{table}[t]
\footnotesize
\caption{Chain-strategy ablation: detection recall on ground-truth APIs when each strategy is individually disabled. IDOR counts ground-truth vulnerable endpoints (5); UAF counts the confirmed endpoint on VVSM-API; Injection counts confirmed vulnerability categories (3: OS~Cmd, XSS, SQLi).}
\label{tab:chain_ablation}
\centering
\begin{tabular}{lccc}
\toprule
\textbf{Variant} & \textbf{IDOR} & \textbf{UAF} & \textbf{Injection} \\
                 & \textbf{(5 endpoints)} & \textbf{(1 schema)} & \textbf{(3 categories)} \\
\midrule
\gq{} (full)       & 5/5 & 1/1 & 3/3 \\
\gq{}-noIDOR       & 0/5 & 1/1 & 3/3 \\
\gq{}-noUAF        & 5/5 & 0/1 & 3/3 \\
\gq{}-noInj        & 5/5 & 1/1 & 0/3 \\
\bottomrule
\end{tabular}
\end{table}

The ablation confirms that each strategy contributes exclusively to its target vulnerability class. Disabling the IDOR strategy eliminates all 5 IDOR detections while leaving UAF and injection results unchanged. Disabling UAF synthesis removes the UAF detection on VVSM-API. Disabling injection detectors drops all three injection categories. This orthogonality is a direct consequence of the modular \texttt{BaseChainStrategy} design: strategies share the dependency graph but execute independent detection logic. One noteworthy cross-API finding: IDOR-API, which implements soft-delete semantics on \texttt{deleteNote} and \texttt{deleteOrder}, also triggers the UAF chain---after \texttt{createNote}~$\to$~\texttt{deleteNote}, the post-delete \texttt{getNote} step returns the deleted note with HTTP~200 and full data. This serendipitous detection is not counted in the formal UAF recall (1/1 covers VVSM-API only); it confirms that chain strategies generalize beyond their primary target APIs and can surface unexpected vulnerability combinations not present in the original ground-truth labeling.

Within these controlled benchmarks, each class-specific detection component is necessary for its vulnerability class: disabling it removes all corresponding detections while leaving the other reported classes unchanged. This experiment does not independently ablate the topological traversal strategy. The full \gq{} configuration is required to detect all three classes simultaneously, addressing {\bf RQ3}.

\subsection{BenGQL Reproducible Benchmark}
\label{sec:bengql}

To complement the live-API results, we run \gq{} and \textsf{EvoMaster} through BenGQL's fixed, Dockerized case studies~\cite{bengql2025}, using a 3{,}600\,s per-tool budget on one machine. BenGQL has no \textsf{ZAP} or \textsf{BurpSuite} runner. Its positive-coverage metric accepts any error-free response, whereas Table~\ref{tab:coverage_table} additionally requires non-null data. The harness also carries a DVGA case study, which we exclude: that run terminated after 5.8\,s with all 19 operations returning errors, yielding no usable coverage metrics, and \textsf{EvoMaster} has no corresponding run. DVGA is instead evaluated directly in Section~\ref{sec:chain-vuln-eval}, where a full run reaches 13/19 operations. Results are in Table~\ref{tab:bengql}.

\begin{table}[t]
\footnotesize
\caption{BenGQL benchmark results (3{,}600\,s budget). \textit{GL(+)}: \gq{} positive coverage. \textit{GL(-)}: \gq{} negative coverage. \textit{EM(+)}: \textsf{EvoMaster} positive coverage. \textit{EM(-)}: \textsf{EvoMaster} negative coverage. `---': the \textsf{EvoMaster} run terminated before emitting its endpoint summary, so no coverage metrics are available for it.}
\label{tab:bengql}
\centering
\begin{tabular}{lcccc}
\toprule
\textbf{Case Study} & \textbf{GL(+)} & \textbf{GL(-)} & \textbf{EM(+)} & \textbf{EM(-)} \\
\midrule
Countries           & 6/6 (100\%)   & 0/6 (0\%)    & 6/6 (100\%)   & 0/6 (0\%)   \\
Fruits-API          & 7/7 (100\%)   & 3/7 (43\%)   & 5/7 (71\%)    & 2/7 (29\%)  \\
Rick \& Morty       & 9/9 (100\%)   & 6/9 (67\%)   & 9/9 (100\%)   & 0/9 (0\%)   \\
EHRI-rest           & 19/19 (100\%) & 9/19 (47\%)  & 19/19 (100\%) & 0/19 (0\%)  \\
React-Finland       & 13/13 (100\%) & 4/13 (31\%)  & 7/13 (54\%)   & 6/13 (46\%) \\
EMB-NCS             & 6/6 (100\%)   & 1/6 (17\%)   & ---            & ---          \\
EMB-SCS             & 11/11 (100\%) & 0/11 (0\%)   & ---            & ---          \\
\midrule
\textbf{Total} & \textbf{71/71 (100\%)} & \textbf{23/71 (32\%)} & \textbf{46/54 (85\%)} & \textbf{8/54 (15\%)} \\
\bottomrule
\end{tabular}
\end{table}

\gq{} reaches 71/71 (100\%) positive coverage across the seven completed case studies. \textsf{EvoMaster} produced no endpoint summary for EMB-NCS or EMB-SCS, where its runs terminated at 32\% and 59\% of the search budget, so both are excluded from its totals. On the remaining five, \gq{} reaches 54/54 (100\%) against \textsf{EvoMaster}'s 46/54 (85\%).

\section{Discussion}
\label{sec:discussion}

\noindent\textbf{Generalizability.}
Any vulnerability whose precondition depends on prior API state is a candidate for a chain strategy. Broken Function Level Authorization (BFLA) can be framed as a two-step chain; race-condition vulnerabilities require concurrent chains. Each strategy is a subclass of a common \texttt{ChainStrategy} interface, so new vulnerability classes can be added without modifying the core engine.

\noindent\textbf{Scalability and Extensibility.}
Two costs dominate compilation. Edge inference compares each of the $F$ \texttt{ID}-typed input fields against all $T$ declared object types, so it is $O(F \cdot T \cdot \ell)$, where $\ell$ bounds the cost of one name comparison. Given the graph, the SCC condensation that orders traversal is $O(|V|+|E|)$ via Tarjan's algorithm~\cite{4569669}. Empirically the combination is inexpensive: Universe, at 176 operations one of the larger schemas we evaluate, compiles in under two seconds. New injection categories plug into the detector registry, and new vulnerability classes subclass \texttt{BaseChainStrategy}, without modifying core fuzzing logic.

\noindent\textbf{Heuristic vs.\ LLM Dependency Resolution.}
\label{sec:llm-vs-heuristic}
\gq{}'s dependency graph is constructed by either the lightweight \emph{heuristic} resolver of Section~\ref{sec:methodology}---name-pattern derivation followed by Levenshtein matching---or an optional \emph{LLM-backed} resolver that reasons over the full schema context.
We compare both resolvers on 10 APIs using \textsf{claude-3-5-sonnet-20241022} via LiteLLM: eight RQ1 APIs plus HIVDB~\cite{hivdb}, the Stanford HIV Drug Resistance Database's public API, and API~Security~\cite{apiSecurityApi}, a small user-and-post service from \gq{}'s sample-API suite; JSON-GraphQL, PokeAPI, and Universe are excluded. This comparison uses a stronger model than the \texttt{gpt-4o-mini} default of Section~\ref{sec:implementation}, so it characterizes the resolver gap rather than the default configuration's behaviour. The 227-operation corpus contains six mutation-bearing and four query-only APIs; its 26-operation Food Delivery snapshot predates the 10-operation RQ1 snapshot. RQ1--RQ3 use the heuristic resolver.

The resolvers agree on 173/227 operations (76.2\%), and the 46 disagreements on mutation-bearing APIs follow three patterns. First, the LLM resolves ambiguous field names such as \texttt{payerID}~$\to$~\texttt{User} that the heuristic labels \texttt{UNKNOWN}, which accounts for most of the 67 LLM-only edges in Table~\ref{tab:resolver_comparison}. Second, the heuristic maps plural field names to non-existent types where the LLM correctly singularizes, which accounts for the 14 heuristic-only edges. Third, nine AniList toggle mutations are labeled \texttt{CREATE} by the heuristic but \texttt{UPDATE} by the LLM, the semantically correct assignment. Agreement reaches 92--100\% on standard-named APIs but falls to 37.5\% on AniList, where the LLM recovers many soft dependencies.

\begin{table}[t]
\footnotesize
\caption{Resolver agreement. \emph{Ops}: operations; \emph{LLM+}/\emph{Heur+}: edges added by only that resolver.}
\label{tab:resolver_comparison}
\centering
\resizebox{\columnwidth}{!}{%
\begin{tabular}{lrrrrrr}
\toprule
\textbf{API} & \textbf{Ops} & \textbf{Agree} & \textbf{Differ} & \textbf{Rate} & \textbf{LLM+} & \textbf{Heur+} \\
\midrule
GraphQLZero     & 32 & 30 &  2 & 93.8\% & 0 & 2 \\
User Wallet  & 26 & 24 &  2 & 92.3\% & 2 & 2 \\
Food Delivery& 26 & 24 &  2 & 92.3\% & 2 & 2 \\
AniList      & 56 & 21 & 35 & 37.5\% &55 & 0 \\
HIVDB        & 41 & 36 &  5 & 87.8\% & 0 & 0 \\
API Security &  6 &  6 &  0 &100.0\% & 0 & 0 \\
Countries    &  6 &  3 &  3 & 50.0\% & 3 & 3 \\
Rick \& Morty&  9 &  6 &  3 & 66.7\% & 3 & 3 \\
EHRI         & 19 & 19 &  0 &100.0\% & 0 & 0 \\
TCGdex       &  6 &  4 &  2 & 66.7\% & 2 & 2 \\
\midrule
\textbf{Total} & \textbf{227} & \textbf{173} & \textbf{54} & \textbf{76.2\%} & \textbf{67} & \textbf{14} \\
\bottomrule
\multicolumn{7}{l}{\footnotesize Overall agreement $p < 10^{-15}$ (binomial, $H_0$: rate $\leq$ 0.5). HIVDB 5 type-only diffs.} \\
\end{tabular}
}
\end{table}

\noindent\textbf{DELETE-Based IDOR.}
\gq{} detects DELETE-based IDOR by replaying the DELETE operation under the secondary credential and flagging a \texttt{\{"ok":~true\}} (HTTP~200) response as evidence that the unprivileged user successfully deleted a resource they do not own. On IDOR-API, both \texttt{deleteNote} and \texttt{deleteOrder} are confirmed under this pattern. Chains where the secondary-user DELETE returns HTTP~0 or an error are not flagged, preventing false positives from legitimate ownership enforcement.

\noindent\textbf{Limitations.}
The injection suite has known blind spots: the path-traversal detector flags payload reflection but does not confirm file-content leakage; time-based blind SQLi requires manual delay calibration; second-order injection is not modeled. Object-type nodes in the dependency graph can appear as IDOR chain candidates, generating empty-response alerts that require automated filtering. Coverage metrics measure reachability and error induction rather than confirmed exploitation---the same rationale used in REST API testing~\cite{restler,resttestgen}. When introspection is disabled, \gq{} falls back to a field-name dictionary, producing incomplete graphs; traffic-based schema reconstruction is future work. Scalar fields with no schema-level type information (e.g., free-form \texttt{String} arguments such as \texttt{family} or \texttt{origin}) receive randomly generated values; when a mutation requires multiple such fields to satisfy server-side validation, the mutation may consistently fail, leaving dependent queries unreachable. Evolutionary search tools can overcome this through adaptive payload refinement---a direction for future work in \gq{}'s scalar materialization.

\section{Ethical Considerations}
\label{sec:ethical}
All testing on public and platform APIs was conducted within the scope of each API's publicly documented terms of service; no credentials were obtained by deception, and no data was exfiltrated beyond what was necessary to compute coverage metrics. Potential vulnerabilities were privately disclosed to the respective vendors before submission, following coordinated disclosure norms. The three ground-truth APIs (IDOR-API, Injection-API, VVSM-API) were designed and deployed by the research team on isolated infrastructure; no real user data was involved. The FinServ API was tested under a formal research agreement with the institution; findings were shared with their security team prior to submission.

\section{Threats to Validity}
\label{sec:threats}

\noindent\textbf{Internal Validity.} Each API--tool pair was scheduled for ten runs, but failures and remote-service availability reduced completion for some pairs; Table~\ref{tab:coverage_table} averages completed runs, and the archive records counts. EvoMaster yielded only one usable AniList and PokeAPI run and two on EHRI, and exhausted memory on GitLab, limiting those estimates. Its 3{,}600\,s budget matches \gq{}'s per-session cap. The research team designed the three ground-truth APIs, so their results are controlled capability demonstrations rather than prevalence estimates. Two coverage-set APIs, {\pt User Wallet}~\cite{UserWalletApi} and {\pt Food Delivery}~\cite{Li_Sample_Food_Delivery_2023}, were also authored by members of the team; the remaining nine are third-party. Component ablations reduce detector coupling; DVGA provides an independent injection oracle, but not independent IDOR or UAF validation.

\noindent\textbf{External Validity.} The 21 APIs vary widely in size, but chain evidence centers on three small purpose-built APIs. The incidental IDOR-API UAF observation is only within-suite cross-schema evidence, and DVGA validates injection on one external application. Larger schemas, non-standard authentication, disabled introspection, and different vulnerability semantics may behave differently.

\noindent\textbf{Conclusion Validity.} RQ1 uses observable black-box coverage and averages completed runs; ground-truth APIs add precision and recall. Per-API results and completion counts support independent interpretation.

\section{Conclusion}
\label{sec:conclusion}

\gq{} turns schema dependencies into executable security preconditions. On the 11 public APIs of the coverage set, it achieves 85.52\% mean \textit{PositiveCoverage}, versus 29.29\% for \textsf{EvoMaster} and 21.80\% for \textsf{ZAP}. Beyond reachability, across the 21 evaluated APIs \gq{} detects all 5 confirmed IDOR endpoints, the formal UAF vulnerability on VVSM-API, incidental post-delete behavior on IDOR-API, and three injection categories. DVGA independently validates XSS and SQLi detection on a third-party application, while every baseline detects zero chain-based vulnerabilities.

These results answer RQ1 with broader operation reachability and RQ2 with controlled evidence that dependency-aware chains expose vulnerabilities missed by single-operation tools. The RQ3 ablation shows that disabling IDOR, UAF, or injection detection removes findings only in its target class. IDOR-API, Injection-API, and VVSM-API are released as open labeled benchmarks, addressing the absence of ground-truth GraphQL chain-vulnerability targets. Independent external validation of IDOR and UAF remains future work.

\section{Data-Availability Statement}
\label{sec:data}

All data supporting the results of this paper---including API coverage measurements, benchmark results, and vulnerability detection comparisons---are archived on Zenodo~\cite{graphqler-artifact}. \gq{} itself is open-source at \url{https://github.com/omar2535/GraphQLer}; the example GraphQL APIs used in this work are available at \url{https://github.com/omar2535/GraphQLer/tree/main/sample-graphql-apis}.

\balance
\bibliographystyle{ACM-Reference-Format}
\bibliography{references}

\end{document}